\documentclass[amsmath,superscriptaddress,reprint,aps]{revtex4-2}
\usepackage{graphicx}
\usepackage{dcolumn}
\usepackage{bm}
\usepackage{hyperref}
\usepackage{siunitx}
\usepackage{color, soul, ulem}

\setstcolor{red}

\begin{document}
	
	
	\title{Centi-combs: Low-noise sub-GHz repetition-rate soliton frequency combs from crystalline resonators}

      \author{Tatsuki~Murakami}
      	\affiliation{Department of Physics, Faculty of Science and Technology, Keio University, Yokohama, 223-8522, Japan}

         \author{Keisuke~Ogawa}
	\affiliation{Department of Physics, Faculty of Science and Technology, Keio University, Yokohama, 223-8522, Japan}
 
     \author{Hajime~Kumazaki}
	\affiliation{Department of Physics, Faculty of Science and Technology, Keio University, Yokohama, 223-8522, Japan}
    
	\author{Shun~Fujii}
	\email[Corresponding author. ]{shun.fujii@phys.keio.ac.jp}
 	\affiliation{Department of Physics, Faculty of Science and Technology, Keio University, Yokohama, 223-8522, Japan}
	
	
\begin{abstract}
We demonstrate low-noise Kerr soliton frequency combs with repetition rates below 1~GHz in ultrahigh-Q crystalline magnesium fluoride resonators. Single soliton states with repetition rates of 0.90~GHz, 1.19~GHz, 1.59~GHz, 2.48~GHz, and 4.10~GHz are observed with continuous-wave laser excitation. The near-GHz soliton repetition frequency exhibits a single-sideband phase noise of -137~dBc/Hz at a 100~kHz offset, surpassing state-of-the-art microwave generators. These ``centi-combs'' bridge the gap between conventional mode-locked lasers and microresonator frequency combs, providing a new route towards real-time sampling, optical-to-microwave synchronization, and hybrid optical clock networks in a compact form. This work expands the operational range of Kerr soliton microcombs from the terahertz to the sub-gigahertz domain, opening new frontiers for frequency comb technologies.

\end{abstract}

	\maketitle
	

The advent of microresonator frequency combs has revolutionized optical frequency comb technologies since the first report in 2007~\cite{Del’Haye2007}. It began with a monolithic silica microresonator with a free-spectral range (FSR) exceeding 1~THz~\cite{Del’Haye2007,DelHaye2011}, and a few years later, a mode-locked state was discovered in a crystalline microresonator with a repetition rate of $\sim$35~GHz~\cite{Herr2014}. The recent development of nanofabrication techniques based on silicon photonics has progressively accelerated integrated photonic platforms, including silicon nitride~\cite{Liu2020,Ye2023}, thin-film lithium niobate~\cite{Zhang2019,Li2023}, III--V aluminum gallium arsenide~\cite{Chang2020}, and various novel materials and combinations~\cite{Kovach2020}. Now, microresonator frequency combs (microcombs) are ubiquitous in a wide range of applications~\cite{Sun2023}, including precision spectroscopy~\cite{Suh600,Stern2020}, ranging~\cite{Suh884,Trocha887}, convolutional processing~\cite{Feldmann2021}, communications~\cite{Marin-Palomo2017,Fueloep2018}, micro- and terahertz-wave generation~\cite{Liang2015:high,Liu2020NP,Tetsumoto2021}, frequency synthesis~\cite{Spencer2018}, and time standards~\cite{Papp2014}.

A high repetition rate, namely a short round-trip length, is central to microcomb applications, as the individual comb lines are spectrally resolved and allow direct access to high-frequency microwave and terahertz regions without electrical multiplication. Nevertheless, low-repetition-rate microcombs, ranging from sub-gigahertz (GHz) to a few gigahertz, remain essential for specific fields and applications such as coherent link between photonics and electronics~\cite{Wu2025}, real-time imaging~\cite{Bao2019a}, signal processing, and clock synchronization~\cite{Zhang2025}.

However, owing to the unavoidable scaling law regarding the threshold of nonlinear parametric oscillation, $\sim V/Q^2$, ultrahigh-Q factors are necessary to compensate for the large mode volume ($V$) in order to generate coherent frequency combs, which is still not an easy task for integrated platforms due to the fundamental limitations of optical loss. In 2018, Suh et al. reported soliton microcombs with repetition rates as low as 1.86~GHz in silica wedge disk resonators with an intrinsic Q of $4.6\times10^8$~\cite{Suh:18}. Such highly optimized nanofabrication procedures enable high-quality microresonators on a chip, but the fabrication of resonators with diameters of several tens of millimeters using conventional lithography still requires further effort to achieve drastic improvements.

\begin{figure}[b!]
\centering\includegraphics{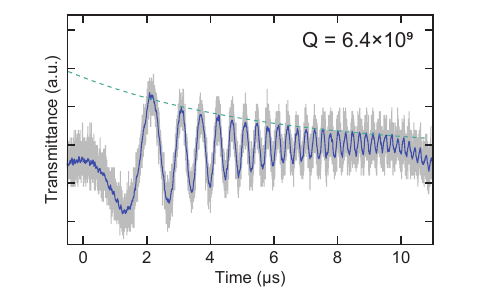}
\caption{\label{Fig_Q} Cavity ring-down measurement of a 900~MHz-FSR resonator, yielding an ultrahigh Q factor of $6.4\times10^9$. The data (gray), moving average (blue), and exponential fit (green) are shown.}
\end{figure}

A pulsed driving technique allows for low-repetition-rate coherent frequency combs~\cite{Xu2020}, and meanwhile, researchers have pursued GHz-rate repetition frequencies in mode-locked laser systems by adopting integrated saturable absorbers~\cite{Martinez2011,Dai2025}, harmonic mode-locking~\cite{Lee2025}, and Kerr-lens configurations~\cite{Kimura2019}. Nonetheless, Kerr soliton microcombs are the prime choice because of their ultrashort pulse widths, intrinsic low-noise properties, and an ultimately simple configuration.


\begin{figure*}[t]
\centering\includegraphics{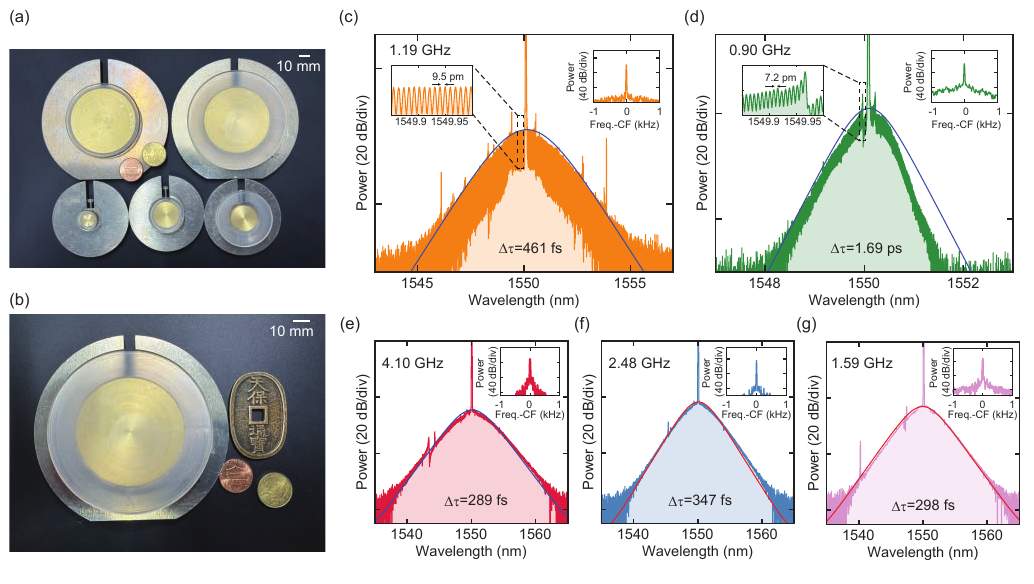}
\caption{\label{Fig_spec} (a,b) Top-view photographs of the fabricated $\mathrm{MgF_2}$ crystalline resonators, shown alongside a U.S. dime, a euro coin, and a Japanese Tenp\={o}-Ts\={u}h\={o} coin. Five resonators with diameters of 58~mm, 77~mm, 44~mm, 28~mm, and 17~mm are arranged in a clockwise direction, and the largest resonator yielding a sub-GHz soliton comb. (c–g) Measured optical spectra of single-soliton combs with $\sim\mathrm{sech^2}$ fitting curves. The insets show the electrical spectra of the soliton beat note with a resolution bandwidth of 10~Hz. $\Delta \tau$ denotes the pulse duration. CF: center frequency.}
\end{figure*}

Here, we report soliton microcombs with repetition rates below 1~GHz in ultrahigh-Q crystalline resonators. Gigantic whispering-gallery-mode (WGM) resonators boasting ultra-low-loss properties allow repetition rates as low as 900~MHz, marking the lowest repetition rate in ``micro''-combs to date. We also demonstrate single-soliton states with repetition rates of 1.19, 1.59, 2.48, and 4.10~GHz. This work opens up a new avenue for optical frequency comb technologies, covering repetition rates from the megahertz to terahertz regime in a single monolithic resonator.

Single-crystal magnesium fluoride ($\mathrm{MgF_2}$) resonators are fabricated by machining and polishing $z$-cut bulk crystals with a thickness of 2~mm. The diameters are roughly adjusted by an initial cylindrical cutting process, and subsequent polishing yields Q factors exceeding several billion, as shown in Fig.~\ref{Fig_Q}. For soliton comb generation, the resonators are pumped by a narrow-linewidth, continuous-wave (CW) fiber laser at 1550~nm via a tapered fiber after light amplification with an erbium-doped fiber amplifier (EDFA), where the maximum pump power is approximately 1.5~W. The detuning locking via the sideband Pound–Drever–Hall (PDH) method is employed to actively capture the soliton state~\cite{Fujii2023,PhysRevLett.122.013902}.

\begin{figure}[h!]
\centering\includegraphics{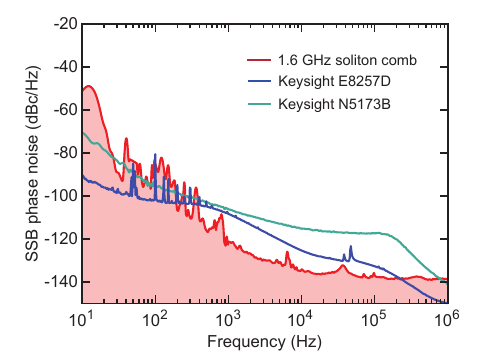}
\caption{\label{Fig_PN} SSB phase noise of a 1.6~GHz soliton comb, compared with commercial benchtop electronic microwave signal generators.}
\end{figure}

Figure~\ref{Fig_spec}(a) and \ref{Fig_spec}(b) show photographs of the fabricated resonators. We fabricated five resonators of different sizes, which yield soliton repetition rates of 4.10~GHz, 2.47~GHz, 1.59~GHz, 1.19~GHz, and 0.90~GHz. The optical spectra and soliton beat notes are presented in Figs.~\ref{Fig_spec}(c)–\ref{Fig_spec}(g). The comb output after rejecting the residual pump power is detected by a photodetector and monitored with a spectrum analyzer. The soliton combs are reproducibly observed in the experiments, and no sophisticated dispersion engineering is performed during fabrication, as $\mathrm{MgF_2}$ resonators naturally exhibit anomalous dispersion at 1550~nm~\cite{Fujii2020,Fujii:20}. Multi-soliton states predominantly appear during the experiment, whereas careful coupling adjustment and backward tuning~\cite{Guo2017} allow the selective excitation of single-soliton states. For repetition rates below 1.2~GHz, an ultrahigh-resolution optical spectrum analyzer (Yokogawa, AQ6380) with a resolution of 5~pm is used to resolve the comb lines and confirm single-soliton operation.

To confirm the noise characteristics of the GHz-repetition-rate ``centi''-comb, the single-sideband (SSB) phase noise is measured through optoelectronic conversion using a photodetector. The representative result is shown in Fig.~\ref{Fig_PN}, where the 1.6~GHz single-soliton state is compared with benchtop electronic microwave generators. The measured SSB phase noise of the soliton comb is -137~dBc/Hz at a 100~kHz offset (-139~dBc/Hz at a 1~MHz offset) and shows a nearly flat plateau region at high offset frequencies, substantially surpassing that of other high-performance electronic oscillators. The soliton phase noise shows a crossover around an offset of 200~Hz, mainly due to the thermal instability of the resonator, resulting in fluctuations of the repetition rate at low-offset frequency regions. The phase noise in this region can be disciplined by microwave injection locking~\cite{Papp2014,PhysRevLett.122.013902} or by sophisticated feedback control~\cite{Kwon2022,Lucas2020,Fujii2024}.

Crystalline materials have shown several advantages over other resonator materials owing to their low thermo-optic coefficient, $n_T$, which is key to achieving ultimately low-phase-noise photonic oscillators~\cite{Liang2015:high,Lucas2020}. Although the thermo-refractive noise scales as $\sim n_T^2 D_1 / A$, where $D_1$ is the FSR and $A$ is the effective mode area, the soliton threshold power increases with the mode volume, $V \approx A / D_1$~\cite{Yao2022}. Therefore, ultrahigh-Q crystalline resonators are indispensable for attaining low-phase-noise soliton combs with gigahertz repetition rates.

In conclusion, we have demonstrated low-noise, low-repetition-rate soliton microcombs in monolithic crystalline resonators. The repetition rate of the microcombs reaches the sub-GHz frequency band for the first time, to the best of our knowledge. This work paves the way for new frontiers in optical frequency comb technologies, bridging conventional mode-locked laser combs and chip-integrated microcombs, and advancing next-generation applications such as timekeeping, phase synchronization with optical lattice clocks, and ultralow-jitter clock links with electronic circuits.

\section*{Funding}
JST PRESTO (JPMJPR25L9); JST Adaptable and Seamless Technology transfer Program through Target-driven R\&D (A-STEP)  (JPMJTR23RF); JSPS KAKENHI (JP24K17624); Keio University Program for the Advancement of Next Generation Research Projects.

\section*{Acknowledgments}
The authors thank Atsushi Ishizawa at Nihon University for providing reference phase noise data.

	\bibliography{Soliton_202510}
\end{document}